\documentstyle[12pt,amsfonts]{article}
\parskip=\medskipamount

\def\?{\hglue1\parindent\ignorespaces} 

\newcommand {\cH}{{\cal H}}

\newcommand {\cK}{{\cal K}}
\newcommand {\cL}{{\cal L}}
\newcommand {\cM}{{\cal M}}

\newcommand {\cO}{{\cal O}}
\newcommand {\cP}{{\cal P}}

\newcommand{\bC}{{\bf C}}

\newcommand{\bZ}{{\bf Z}}
\def\a{\alpha}

\def\d{\delta}

\def\G{\Gamma}

\def\m{\mu}

\def\s{\sigma}

\def\D{\Delta}

\def\F{\Phi}
\def\J{\Psi}
\def\L{\Lambda}

\def\S{\Sigma}
\def\U{\Upsilon}
\def\X{\Xi}

\newcommand{\pa}{\partial}                           
\newcommand{\hf}{\frac12}

\newcommand{\sect}[1]{\setcounter{equation}{0}\section{\boldmath#1}}

\newcommand{\be}{\begin{equation}}
\newcommand{\ee}{\end{equation}}
\newcommand{\bea}{\begin{eqnarray}}
\newcommand{\eea}{\end{eqnarray}}
\newcommand{\non}{\nonumber}

\def\Bar#1{\overline{#1}}
\newcommand{\beq}{\begin{equation}}
\newcommand{\eeq}{\end{equation}}

\def\fracm#1#2{\hbox{\large{${\frac{{#1}}{{#2}}}$}}}

\font\ro=cmsy10       
\def\kcr{{\hbox{\ro \char'170}}}    
\def\ktl{{\hbox{\ro \char'170}}}  
\def\ktr{{\hbox{\ro \char'170}}}  
\def\kbl{{\hbox{\ro \char'170}}}  
\def\kbr{{\hbox{\ro \char'170}}}  

\def\endtitle{\end{quotation}\newpage}     

\def\headpic{           
  \indent
  \setlength{\unitlength}{.4mm}
  \thinlines
  \par
  \begin{picture}(29,16)
  \put(165,16){\line(1,0){4}}
  \put(170,16){\line(1,0){4}}
  \put(180,16){\line(1,0){4}}
  \put(175,0){\line(1,0){4}}
  \put(180,0){\line(1,0){4}}
  \put(185,0){\line(1,0){4}}
  \put(169,0){\line(0,1){16}}
  \put(170,0){\line(0,1){16}}
  \put(179,0){\line(0,1){16}}
  \put(180,0){\line(0,1){16}}
  \put(184,0){\line(0,1){16}}
  \put(185,0){\line(0,1){16}}
  \put(169,16){\oval(8,32)[bl]}
  \put(170,16){\oval(8,32)[br]}
  \put(179,0){\oval(8,32)[tl]}
  \put(185,0){\oval(8,32)[tr]}
  \end{picture}
  \par\vskip-6.5mm
  \thicklines}

\def\border{           
  \setlength{\unitlength}{1mm}
  \newcount\xco
  \newcount\yco
  \xco=-21
  \yco=12
  \begin{picture}(140,0)
  \put(\xco,\yco){$\ktl$}
  \advance\yco by-1
  {\loop
  \put(\xco,\yco){$\kcr$}
  \advance\yco by-2
  \ifnum\yco>-240
  \repeat
  \put(\xco,\yco){$\kbl$}}
  \xco=158
  \yco=12
  \put(\xco,\yco){$\ktr$}
  \advance\yco by-1
  {\loop
  \put(\xco,\yco){$\kcr$}
  \advance\yco by-2
  \ifnum\yco>-240
  \repeat
  \put(\xco,\yco){$\kbr$}}
  \put(-20,13){\tiny University of Maryland Elementary Particle
Physics University of Maryland Elementary Particle Physics University of
Maryland Elementary Particle Physics}
  \put(-20,-241.5){\tiny University of Maryland Elementary
Particle Physics University of Maryland Elementary Particle Physics
University of Maryland Elementary Particle Physics}
  \end{picture}
  \par\vskip-8mm}

\catcode`@=11    
\let\down=\downarrow
\def\crlap#1{\rlap{$\vcenter{\hbox{$\scriptstyle#1$}}$}}
\def\cllap#1{\llap{$\vcenter{\hbox{$\scriptstyle#1$}}$}}
\newbox\t@b@x
\def\rightarrowfill{$\m@th \mathord- \mkern-6mu
     \cleaders\hbox{$\mkern-2mu \mathord- \mkern-2mu$}\hfill
      \mkern-6mu \mathord\rightarrow$}
\def\leftarrowfill{$\m@th \mathord\leftarrow \mkern-6mu
     \cleaders\hbox{$\mkern-2mu \mathord- \mkern-2mu$}\hfill
      \mkern-6mu \mathord-$}
\def\leftrightarrowfill{$\m@th \mathord\leftarrow \mkern-6mu
     \cleaders\hbox{$\mkern-2mu \mathord- \mkern-2mu$}\hfill
      \mkern-6mu \mathord\rightarrow$}
\def\tooo#1{\setbox\t@b@x=\hbox{\footnotesize$#1$}%
             \mathrel{\mathop{\hbox to\wd\t@b@x{\rightarrowfill}}%
              \limits^{#1}}\,}
\def\froto#1{\setbox\t@b@x=\hbox{\footnotesize$#1$}%
             \mathrel{\mathop{\hbox to\wd\t@b@x{\leftrightarrowfill}}%
              \limits^{#1}}\,}
\catcode`@=12                   

\begin{document}
\border\headpic {\hbox to\hsize{February 1999 \hfill {UMDEPP 99-99}}}
\par
\setlength{\oddsidemargin}{0.3in}
\setlength{\evensidemargin}{-0.3in}
\begin{center}
\vglue .03in

{\large\bf\boldmath CNM Models, Holomorphic Functions and \\
Projective Superspace $C$-Maps}\footnote{Supported in
part by US NSF Grant PHY-98-02551,
the US DOE Grant DE-FG02-94ER-40854,\\ \?\?
and the Deutsche Forschungsgemeinschaft.}
\\[.3in]
S.~James Gates, Jr.\\[-1mm]
{\it Department of Physics,
University of Maryland at College Park \\[-1mm]
College Park, MD 20742-4111, USA\/}\\[-1mm]
{\tt gates@bouchet.physics.umd.edu}\\[0.2in]

Tristan H\"ubsch \footnote{On leave from  the
``Rudjer Bo\v skovi\'c'' Institute, Zagreb, Croatia.}\\[-1mm]
{\it Department of Physics and Astronomy\\[-1mm]
Howard University, Washington, DC 20059, USA\/}\\[-1mm]
{\tt thubsch@howard.edu}\\[0.1in]
and
\\[0.1in]

Sergei M.~Kuzenko \footnote{On leave from  the
Department  of Physics, Tomsk State University, Russia} \\ [-1mm]
{\it Institut f\"ur Theoretische Physik,  Universit\"at
M\"unchen\\[-1mm]
Theresienstr. 37, D-80333 M\"unchen, Germany\/}\\[-1mm]
{\tt Sergei.Kuzenko@Physik.Uni-Muenchen.DE}\\[0.2in]

{\bf ABSTRACT}\\[.002in]
\end{center}
\begin{quotation}
{Continuing the investigation of CNM (chiral-nonminimal) hypermultiplet
nonlinear $\s$-models, we propose extensions of the concept of the
$c$-map which relate holomorphic functions to hyper-K\"ahler geometries.
In particular, we show that a whole series of hyper-K\"ahler potentials
can be derived by replacing the role of the 4D, $N{\,=\,}1$ tensor
multiplet in the original $c$-map by 4D, $N{\,=\,}1$ non-minimal multiplets
and auxiliary superfields. The resulting $N{\,=\,}2$ models appear to have
interesting connections to Calabi-Yau manifolds and algebraic varieties. 
These models also emphasize the fact that special hyper-K\" ahler
manifolds (the analogs of special K\" ahler manifolds) without
isometries exist.}\endtitle

\sect{Introduction}
\? The study of supersymmetrical nonlinear $\s$-models has proven
repeatedly to provide a way in which to derive new results in complex
manifold theories with potentials.  One such example of this is
provided by the ``$c$-map,'' known for approximately a decade due to
a work of Cecotti, Ferrara and Girardello~\cite{CMAP}.   According to
the Glossary in \cite{polchinski} the $c$-map is ``a method for constructing
the hypermultiplet  moduli space of a type-II string theory compactified
on a Calabi-Yau  three-fold from the vector multiplet moduli space of the
other type-II theory on the same three-fold''. In {\it rigid
supersymmetry\/}, with which we only concern ourselves in this paper, the
$c$-map  sends a special K\"ahler manifold of the rigid type $\cM_d$ to
a hyper-K\"ahler manifold $\cH_{2d}$ (with both subscripts denoting
the complex dimensions)~\cite{CMAP}.  One feature that arose due
to the method of construction utilized is that the resultant
hyper-K\"ahler metrics always possess a set of isometries.  So clearly
this construction leads to a restricted class of hyper-K\"ahler
geometries.  This naturally raises the question of whether there exist
generalizations of the $c$-map that lead to a wider class of
hyper-K\"ahler geometries.

Recently, we began an effort to investigate the structure of 4D,
$N{\,=\,}2 $ nonlinear $\s$-models which have the property that the
$N{\,=\,}2$ supersymmetry is manifest~\cite{gkCNM} by use of projective
superspace. Projective superspace techniques made their first
appearance in the literature via the work in~\cite{GHR} but received
their most complete development in the later works of~\cite{KLR,ks}.  One
implication of our completed investigation is that
essentially{\it{all}\/}\footnote{There is a mild restriction that in
order to possess an $N{\,=\,}2$ extension, the K\"ahler potential
defining\\ \?\? the $N{\,=\,}1$ nonlinear $\s$-model must  be an analytic
function.} 4D, $N{\,=\,}1$ supersymmetric nonlinear $\s$-models possess
4D, $N{\,=\,}2$ supersymmetric  extensions.  It is thus natural  that we
turn our attention to the question  of whether the projective  approach
can be used to gain more insight into  the issue of generalizations
to the $c$-map.  In this work we will present  results from this effort.
We begin by looking at the classical $c$-map~\cite{ CMAP} and presenting
a streamlined proof as compared to the original one.   Here, using
projective superspace allows the discovery of a simple reason why the
original $c$-map necessarily relates a holomorphic function to a
hyper-K\"ahler  metric.
 We next observe that the original $c$-map is related to a particular
projective multiplet representation, ``the projective O(2)-multiplet.''
This  is followed with the observation that the role of the projective
O(2)-multiplet is not unique and with a simple generalization,
projective O($2n$)-multiplets and finally the projective polar multiplet
may be used  instead.

The structure of our proposed generalizations to the $c$-map suggest
interesting connections to issues in algebraic geometry. In particular
the models which emerge from this approach have the structure of being
theories that are defined on hypersurfaces in K\"ahler spaces.  These
surfaces are defined by holomorphic constraints. An interesting case
for the polar multiplets occurs when these holomorphic constraints are
polynomial in nature.

\sect{The Aboriginal $C$-Map}
\? The work of~\cite{CMAP} started with a holomorphic
prepotential\footnote{Here we use this term to refer to its more recent
definition, not its original one~\cite{GTSG}.} $F(\F)$ defining the
K\"ahler potential $K(\F, \bar \F )$ for a special K\"ahler geometry
with metric $g_{I {\bar  J} } (\F, \bar \F )$
\be
K(\F^I, \bar \F^J ) ~=~ {\bar \F}^I \, F_I (\F) ~+~ \F^I \,
{\bar F}_I (\bar \F) ~~~,~~~ g_{I {\bar J} } (\F, \bar \F )
~=~  F_{IJ}(\F) ~+~ {\bar F}_{IJ}(\bar \F) ~~~.
\label{kahler}
\ee
Next, the following extension of the K\"ahler potential was introduced
\be
H(\F, \bar \F, \J, \bar \J) ~=~ K(\F, \bar \F) ~+~ \fracm 12 g^{I {\bar
J} } (\F, \bar \F) ( \J_I + {\bar \J}_I) ( \J_J + {\bar \J}_J) ~~~.
\label{cmap}
\ee
In~(\ref{kahler}) and~(\ref{cmap}) the quantities $\Phi^I$, as usual,
are chiral superfields that have the geometrical interpretation of being
the coordinates of the complex manifold whose metric in given by $g_{
I {\bar J} }$. In~(\ref{cmap}) the quantities $\J_I$ are also chiral
superfields.  However, their geometrical interpretation is very
different.  Since these appear in the expression multiplying the inverse
K\"ahler metric, it follows that $\J_I$ must be co-vectors associated
with the K\"ahler manifold. After considerable effort it was shown that
$H(\F,\bar \F, \J, \bar \J)$ defines a hyper-K\"ahler geometry.  At this
stage, we see that the $c$-map also has the interpretation of providing a
mechanism for defining hyper-K\"ahler geometries in terms of a purely
holomorphic function $F(\F)$. According to a well-known theorem
regarding supersymmetric non-linear $\s$-models, all such 4D, $N{\,=\,}2$
(or 2D, $N{\,=\,}4$) supersymmetric models describe hyper-K\"ahler
geometries.  An apparent feature of~(\ref{cmap}) is that it is
necessarily invariant  under the change of variables $\J_I \, \to \, \J_I
\, + \,{\rm i} a_I$,  if $a_I$ are a set of real quantities.  This is
equivalent to the appearance of a set of isometries of the hyper-K\"ahler
metric.  So it is an intrinsic feature of the $c$-map (as described
above) to entail a procedure in which some hyper-K\"ahler manifolds that
necessarily contain isometries are  constructed from a single holomorphic
function.

The proof given in~\cite{CMAP} that the potential~(\ref{cmap}) describes
a hyper-K\"ahler geometry is quite involved. Below we will give an
alternate and much more transparent proof of this, taking advantage of 
the fact that any nonlinear 4D $\s$-model that realizes $N{\,=\,}2$
supersymmetry necessarily describes a hyper-K\"ahler geometry.  A set of 
$d$ $N{\,=\,}2$ tensor multiplets is described in $N{\,=\,}1$ superspace
by $d$ chiral superfields $\F^I$ and $d$ real  linear superfields $G^I$.
 
A way to make manifest $N{\,=\,}2$ supersymmetry is to use the projective  
superspace techniques~\cite{KLR} which, in turn, naturally emerge from 
the  fundamental  concept of harmonic superspace \cite{GIKOS}. In this
approach to $N{\,=\,}2$ supersymmetric theories, a whole sequence of scalar
multiplets, known as O($2n$) projective multiplets~\cite{KLR,ks} has been
found.   The case of the O(2) projective multiplet is directly relevant
to the $N{\,=\,}2$ tensor multiplet. The quantity defined by
\be
\X^I (w) ~=~ \F^I ~+~ w G^I ~-~ w^2 {\bar \F}^I ~~~,
  \qquad I=1,\ldots,d~~~, \label{O2M}
\ee
is an O(2) projective multiplet and is allowed to enter a projective
$N{\,=\,}2$ supersymmetric action of the form
\be
S ~=~ \frac{1}{\, 2\pi \,{\rm i} \,} \oint_{C} \frac{{\rm d}w}{w}
\int{\rm d}^8 z\,\,  \cL (\X^I (w) , w)  ~~~,
\label{action}
\ee
where the Lagrangian may be specified to the form
\be
S ~=~ - \frac{1}{\, 2\pi \,{\rm i} \,} \oint_{C} \frac{{\rm d}w}{w}
\int{\rm d}^8 z\,\, \frac{  F(\X^I (w))  }{w^2} ~+~ {\rm h.c.} ~~~,
\label{tensoraction}
\ee
in terms of a single holomorphic function $F$. A simple calculation
gives
\bea
F(\X^I (w)) \,&=&\, F \Big( \F^I + w G^I ~-~ w^2 {\bar \F}^I \Big)
\non \\
&=&\, F(\F) ~+~ w F_I (\F) G^I ~-~ w^2 \Big( F_I (\F) {\bar \F}^I
~-~ \fracm 12 F_{IJ} (\F) \, G^I G^J \Big)
~+~ O(w^3) ~~~. \non
\eea
Therefore, the action is equivalent to
\be
S[\F, \bar \F, G] ~=~ \int{\rm d}^8 z\, \left\{ ~K (\F, {\bar \F})
~-~ \fracm 12 g_{I {\bar J} } (\F, \bar \F ) \, G^I G^J ~ \right\} ~~~.
\label{N2TMact}  \ee
The real linear superfields $G^I $ can be dualized (using an $N{\,=\,}1$
superfield duality transformation) into pairs of (anti)chiral ones,
$\J_I$ and ${\bar \J}_I$. As a result, the action turns into
\be
S[\F, \bar \F, \J, \bar \J ] ~=~ \int{\rm d}^8 z\,\, H(\F, \bar \F, \J,
\bar \J) ~~~,
\label{cmap2}
\ee
with $H(\F, \bar \F, \J, \bar \J)$ given precisely as in~(\ref{cmap}).
A special feature of the potential~(\ref{cmap}) is that the (anti)
chiral superfields $\J$ and $\bar \J$ {\it cannot\/} be converted into
{\it complex\/} (anti)linear ones $\G$ and $\bar \G$!

We thus see in Eqs.~(\ref{O2M}--\ref{cmap2}) a simple and
straightforward
proof of the existence of the $c$-map with the consequence that the
potential~(\ref{cmap}) describes a hyper-K\"ahler geometry.

\sect{Construction of New O($2n$) $C$-Maps}
\? As noted below Eq.~(\ref{cmap2}), the hyper-K\"ahler geometry
there described cannot be obtained from our previous discussion of
CNM-hypermultiplet models~\cite{gkCNM}.  The technical reason for this
is because of the impossibility to perform an $N{\,=\,}1$ superfield
duality transformation from chiral to complex linear superfields
whenever the chiral superfields appear in an action solely via the
linear combination $\J_I + {\bar \J}_I$ and CNM models necessarily
involve complex linear multiplets. This leaves two options:
\begin{itemize}\addtolength{\rightskip}{8.0em plus 2pt}

\item It is impossible to eliminate all the auxiliary superfields
in the CNM-hypermultiplet models appropriate for the case at
hand;

\item If it is possible to eliminate all the auxiliary superfields
in the CNM-hypermultiplet models for K\"ahler potential of
the form~(\ref{kahler}), we apparently arrive at a different
hyper-K\"ahler manifold as compared to that defined by~(\ref{cmap}).
This implies the $c$-map is not unique, at least
in the case of rigid SUSY.

\end{itemize}

Let us repeat the above considerations for the case of O(4) projective
superspace multiplet (instead of the  tensor multiplet). The basic
superfields
read
\be
\S^I (w) ~=~ \F^I + w \G^I +w^2 V^I -w^3 { \bar \G}^I + w^4 {\bar
\F}^I ~~~,~~~ \quad {\bar V}^I ~=~ V^I ~~~, \label{O2Nx}
\ee
where $\F^I$ are chiral, $\G^I$ complex linear, $V^I$ real unconstrained
superfields. Instead of the action~(\ref{tensoraction}), we now have
\be
S_{O(4)} ~=~ \frac{1}{\, 2\pi \,{\rm i}\,} \oint_{C} \frac{{\rm d}w}{w}
\int{\rm d}^8 z\, \frac{ F(\S^I (w)) }{w^4} ~+~ {\rm h.c.} ~~~,
\label{o(4)act}
\ee
or, after performing the contour integral,
\bea
S [\F,\, \bar \F, \,\G, \bar \G,\, V]\, &=& \int{\rm d}^8 z ~ \Big\{
~  K(\F, \, \bar \F ) ~-~  g_{I {\bar J} } (\F, \bar \F )\,  \G^I
{\bar \G}^J ~+~ \fracm 12 g_{I {\bar J} } (\F, \bar \F ) ~ V^I \, V^J
\non {~~~~~~} \\
& {~}&{~~~~~~~~~~~}+~ \fracm 12 \Big[ \,\, F_{IJK} (\F)\, \G^J \G^K
\,+\,
{\bar F}_{IJK} (\bar \F) \, {\bar \G}^J {\bar \G}^K \,\, \Big] \, V^I
\non \\
&{~}&{~~~~~~~~~~~}+~ \fracm{1}{4!} F_{IJKL}\, \G^I \G^J \G^K \G^L ~+~
\fracm{1}{4!}  {\bar F}_{IJKL} \,{\bar \G}^I {\bar \G}^J {\bar \G}^K
{\bar \G}^L ~ \Big\} ~~~ .
\eea
The auxiliary superfields $V^I$ can be easily eliminated via
\bea
V^L ~=~ - \fracm 12 \, g^{I {\bar L} } (\F, \bar \F ) \, \Big[ \,\,
F_{IJK}
(\F)\, \G^J \G^K \,+\,  {\bar F}_{IJK} (\bar \F) \, {\bar \G}^J {\bar
\G}^K \,\, \Big]  ~~~,
\eea
which can be used to re-write the action totally in terms of the chiral
and complex linear superfields.  Then, the action will include terms of
the second- and fourth-order in $\G$ and $\bar \G$. We explicitly find
that the removal of the auxiliary superfields yields,
\bea
S [\F,\, \bar \F, \,\G, \bar \G]\, &=& \int{\rm d}^8 z ~ \Big\{
~  K(\F, \, \bar \F ) ~-~  g_{I {\bar J} } (\F, \bar \F )\,  \G^I
{\bar \G}^J \non {~~~~~~~~~~} \\
& {~}&{~~~~~~~~~~~}-~ \fracm 14 \, F_{MJK} (\F) \, g^{M {\bar N}} \,
{\bar F}_{NRS} (\bar \F) \, \G^J \, \G^K \, {\bar \G}^R \, {\bar \G}^S
\non \\  &{~}&{~~~~~~~~~~~}+~ \fracm{1}{4!} {\cal F}_{IJKL}\, \G^I \G^J
\G^K \G^L ~+~ \fracm{1}{4!}  {\bar {\cal F}}_{IJKL} \,{\bar \G}^I {\bar
\G}^J {\bar \G}^K  {\bar \G}^L ~ \Big\} ~~~ , \non \\
{\cal F}_{IJKL}\, &\equiv& \, F_{IJKL} ~-~ 3 \, F_{IJM}\, g^{M
{\bar N} } \, {F}{}_{KLN} ~~~. \label{O2nact}
\eea
We emphasize that this action despite its relative simplicity 
and written in terms of $N{\,=\,}1$ superfields actually possesses 
$N{\,=\,}2$ supersymmetry.  This is directly analogous to the
original $N{\,=\,}1$ superfield introduction of the $N{\,=\,}2$ 
supersymmetric K\"ahlerian Vector Multiplet (KVM) model~\cite{KVM}.

To obtain the hyper-K\"ahler potential, we have to dualize the
(anti)linear scalars $\G,\bar \G$ into (anti)chiral ones $\J,\bar \J$.
This amounts to implementing an $N{\,=\,}1$ superfield Legendre
transform of $\G$ and $\bar \G$ in the action~(\ref{O2nact}) by writing 
the master action
\be
S^* [\F,\, \bar \F, \,W, \bar W, \J, \,{\bar \J}] ~=~ S [\F,\, \bar \F,
\,W, \bar W] ~+~ \int d^8 z ~ [ ~ W^I\,  \J_I ~+~  \bar W^I \, {\bar
\J}{}_I ~] ~~~,
\ee
and then to determine the unconstrained 
complex superfields $W$ and $\bar W$ 
in terms of $\J$ and $\bar \J$ by solving their equations
of motion
\be
\frac {\d S^*}{\d W^I} ~=~ 0 ~~\to ~~ \J_I ~+~ \frac{\pa{~~\,}} {\pa
W^I}\,
\cL (\F, \bar \F, W, \bar W) ~=~0~~~,
\label{ltr}
\ee
where $ \cL (\F, \bar \F, \G, \bar \G) $ is the Lagrangian
in~(\ref{O2nact}). Since the K\"ahler metric $g_{I {\bar J} } (\F, \bar
\F )$ is nonsingular and taking into account the explicit structure of
$ \cL (\F, \bar \F, \G, \bar \G) $, the above equations~(\ref{ltr})
are solved uniquely in a small neighborhood of the origin in $W$-space. 
Globally, the equations~(\ref{ltr}) are solved exactly  although not
uniquely, see below. The hyper-K\"ahler structure thus derived does not
possess abelian isometries, in contrast to~(\ref{cmap}), obtained via the
original $c$-map.

One of the interesting features of the O($2n$) (for $n \ge  2$) projective
multiplets, in contrast to the O(2) projective multiplet, is that they
can be coupled to Yang-Mills multiplets.  The exception in the case
of the O(2) projective multiplet is due to the fact that only the O(2)
multiplet contains a component level 2-form gauge field. However, the
O($2n$) (for $n \ge$ 2) projective multiplets can only provide
representations of matter that are real representations of the
Yang-Mills gauge group.  This condition arises due to the reality of the
superfield  that occurs as the coefficient of the $w^n$-term in the
O($2n$) projective  multiplet.  Since the action~(\ref{O2nact}) depends
on the holomorphic  function $F$, it is interesting to conjecture that
the O(4) projective $\s$-model can be combined with the superfield
K\"ahlerian Vector Multiplet model~\cite{KVM} by identifying the two {\it
{a}\/} {\it {priori}\/}  independent holomorphic functions as one and the
same.  This seems logical as a step toward the construction of
hyper-K\"ahlerian Vector Multiplet  (HKVM) models. In the limit where $F
= \fracm 12 {\F}^2$ and with a modification of the linearity condition,
the combined KVM action and projective O(4) $\s$-model~(\ref{O2nact})
describes the $N{\,=\,}4$ Yang-Mills model. It  is a topic for additional
study to see if  this $N{\,=\,}4$ supersymmetry continues to exist with
more general choices of $F$ such as the  Seiberg-Witten
prepotential~\cite{SW}.

There is an obvious generalization to the results in~(\ref{O2Nx})
and~(\ref{O2nact}).  Consider the general O($2n$) multiplet of the form
\bea
\S^I (w) \,&=&\, \F^I ~+~ w \, \G^I ~+~
\sum_{\ell =2}^{n-1} w^{\ell} U^I_{\ell -1}
~+~w^{n} \,V^I \non \\
&{~}&\, +~  (-1)^n \left\{
\sum_{\ell =2}^{n-1} (-1)^{\ell}
w^{2n - \ell} {\bar U}^I_{\ell -1}
-w^{2n-1}  \, {\bar \G}^I
~+~ w^{2n} \, {\bar \F}^I
\right\}
~~~, \label{O2nx}
\eea
where $\Phi$ is a chiral superfield, $\G$ is a complex linear
superfield, ${V^I}$ is an arbitrary real general superfield and
the  remaining 4D, $N{\,=\,}1$  superfields $U_{\ell}$ in~(\ref{O2nx})
are complex general ones.   We consider the action of the form
\be
S_{O(2n)} = (-1)^n
\frac{1}{\,2\pi \,{\rm i}\,} \oint_{C} \frac{{\rm d}w}{w} \int{\rm
d}^8
z\,\, \frac{  F(\S^I (w))  }{w^{2n}} ~+~ {\rm h.c.} ~~~. \label{O2NAT}
\ee
This obviously leads to a generalization of~(\ref{o(4)act}) and it
includes a term of the
form
\be
(-1)^n \fracm 12 \, g_{I {\bar J} } (\F, \bar \F ) \,\, {V^I} \, {V^J}
\ee
where the superfield $V^I$ is the real $N{\,=\,}1$ coefficient
superfield of the $w^n$ term in the O($2n$) multiplet $\S^I (w)$.  The
only other manner  in which this superfield appears in~(\ref{O2NAT})
is via a term that is linear  in $V^I$.  That is why $V^I$ can always
be explicitly {\it and uniquely\/} removed by its algebraic equation
of motion.  Clearly, there are lots of other terms involving the
other complex superfields $U^I_{\ell}$ and their conjugates
${\bar U}^I_{\ell}$ for $\ell = 1,...,n-2$. For the full theory we have
\bea
S_{O(2n)} [\F,  \G,  U, V] \,=\, \int{\rm d}^8 z \, &\Big[& K(\F, \bar 
\F ) ~- ~ g_{I {\bar J} } (\F, \bar \F ) \G^I {\bar \G}^J ~+~ {\cal 
P}({U}_{\ell},  {\bar U}_{\ell}, V \,; \, \F, \bar \F,\G , {\bar \G} 
) \non \\
&~&+\,\, (-1)^n \frac{1}{(2n)!}\Big\{ F_{ I_1 I_2 \cdots I_{2n} } 
(\F) \G^{I_1}  \G^{I_2}   \dots  \G^{I_{2n}} ~~+~~{\rm h.c. } \Big\}
\Big] \;, \non \\ \label{HMact}
\eea
where the K\"ahler potential and metric are determined as in~(\ref{kahler}) 
in terms of the holomorphic prepotential in~(\ref{O2NAT}).  The same
holomorphic prepotential $F$ in~(\ref{O2NAT}) also  determines the function
${\cal P}$ being a multinomial in ${U}_{\ell}$,  ${\bar U}_{\ell}$ and $V$
with the coefficients depending polynomially on $\G$, ${\bar \G}$ and
functionally on $\F$, ${\bar \F}$ such that
\bea
{\cal P}( {U},  {\bar U}, V \,; \, \F, \bar \F,\G ,  {\bar \G} )
\Big|_{  {U} =  {\bar U}   = V  = 0} ~&=&~0 \non \\
\frac{\pa^2{~~~~~~}}{ \pa U^I_{\ell} \pa {\bar U}^J_{\ell} }
{\cal P}( {U},  {\bar U}, V \,; \, \F, \bar \F,\G ,  {\bar \G} )
~&=&~
(-1)^{\ell}  g_{I {\bar J} } (\F, \bar \F )
~~~.
\eea
In particular, for the O(6) multiplet model one obtains
\bea
 \cP_{O(6)} &=& g_{I {\bar J} } (\F, \bar \F ) U^I {\bar U}^J
~-~ \fracm 12  g_{I {\bar J} } (\F, \bar \F ) V^I V^J \non \\
&~&
-~ V^I \Big\{ F_{IJK} (\F, \bar \F) \G^J U^K
~+~ \fracm 16 F_{IJKL} (\F, \bar \F) \G^J \G^K \G^L
~+~ {\rm h.c.} ~\Big\} \non \\
&~&-~\fracm 16 \Big\{ F_{IJK} (\F, \bar \F) U^I U^J U^K
~+~ {\rm h.c.} ~\Big\} \non \\
&~&+~ \fracm 12 \Big\{ F_{IJK} (\F, \bar \F) \G^I \G^J {\bar U}^K
-~ \fracm 12   F_{IJKL} (\F, \bar \F) \G^I \G^J U^K U^L ~\non \\
&~~~&~ \qquad  \qquad  -~\fracm {1}{12} F_{IJKLM} (\F, \bar \F)
\G^I \G^J \G^K \G^L U^M  ~+~{\rm h.c.} ~\Big\}~.
\label{O(6)}
\eea

As is already clear from~(\ref{O(6)}),  for $n>2$ we face the
problem of eliminating the auxiliary superfields.
Of course, one can develop a perturbation theory to 
solve the equations of motion for ${U}_{\ell}$, ${\bar U}_{\ell}$
and $V$:
\be
\frac{\pa}{ \pa {U^I}_{\ell}}  {\cal P} ~=~
\frac{\pa}{ \pa {{\bar U}^I}_{\ell}}  {\cal P} ~=~
\frac{\pa }{ \pa {V^I}   }      {\cal P} ~=~0
\label{surface}
\ee
by representing $F(\F) = F_0 +\D F$, where $F_0 = \hf \F^2$
represents the leading contribution to $F (\F)$ and $\D F = \cO (\F^3)$
corresponds to a small perturbation. However, explicit solutions to
equations~(\ref{surface}) can be found only for special choices of
$F(\F)$. In general,  Eqs.~(\ref{surface}) present an algebro-geometric
problem, and we return to this below.

Suffice it here to recall that the coordinate and the vector superfields
$\Phi,\G$ parametrize the total space of the tangent bundle, $T_{\cal M}$,
over the manifold of special K\"ahler geometry whose metric is
determined by the holomorphic function $F$. The auxiliary superfields
$U_\ell, \,V,\,{\bar U}_\ell$ enlarge this space considerably. The
$(2n\,-\,3)\,\, d$ constraint equations~(\ref{surface}), for $n\ge 2$,
then locate $T_{\cal M}$ as an algebraic variety within the enlarged
field space.

For a general holomorphic prepotential $F(\F)$, the O($2n$) multiplets
themselves, $\S^I$~(\ref{O2nx}), possess a  $\bZ_{2n}$ symmetry defined
as follows:
\be
\S^I  (w)~ \longrightarrow ~\S^I  ({\rm e}^{{\rm i} \a} w)
~~~,
~~~ ({\rm e}^{{\rm i}  \a})^{2n} = 1 ~~~.
\label{z2n}
\ee
Here the restriction $\exp (2n {\rm i}\, \a ) =1$ follows from the
twisted reality condition~(\ref{real}) to which the projective
superfields $\S^I (w)$ are subject. It is the requirement of $\bZ_{2n}$
symmetry which forbids us to add  terms of the form
\be
\frac{  F(\S^I (w))  }{w^k}~~, \qquad \qquad k~\neq~0\pmod{2n}~~~,
\ee
to the Lagrangian in~(\ref{O2NAT}).

It should be stressed that for the O(2) model we have not only the
discrete $\bZ_2$ symmetry, but also $d$ abelian noncompact isometries
(Peccei-Quinn symmetry) in addition. The point is the $O(2)$ action
in~(\ref{N2TMact}) is invariant under shifts
\be
G^I (z) ~\longrightarrow ~ G^I (z) ~+~ a^I ~~~,\qquad I=1,\ldots,d~~~,
\ee
with the $a$'s being arbitrary real constants.  Such isometries are
characteristic of the $O(2)$ model only and they cannot
be present for the general O($2n$) projective models with $n>1$.

\sect{Construction of the Minimal Polar Multiplet\protect\newline
$C$-Map}
\? In our previous work~\cite{gkCNM}, we discussed\footnote{Our
notational conventions in this section are those in~\cite{gkCNM}.} the
existence of 4D, $N{\,=\,}2$ nonlinear $\s$-models based on the use
of the polar representation~\cite{GRR} of projective superspace
\bea
\U^I (w) &=&\, \sum_{n=0}^{\infty}  \, \U_n^I (z) w^n ~=~
\F^I(z) + w \G^I(z) + O(w^2) ~~~,\non \\
&\equiv& \, [\,\,  \F^I ~+~ \G^I w  ~+~ {\cal A}^I (w) ~] ~~~, \non \\
\breve{\U}{}^{\bar{I}} (w)  &=&\, \sum_{n=0}^{\infty}  \, {\bar
\U}_n^{\bar
{I}} (z) (\fracm {-1}{w})^n ~=~ {\bar \F}{}^{\bar{I}}(z)  - \fracm 1{w}
{\bar \G}{}^{\bar{I}}(z)  + O((\fracm{1}{w})^2)~~~,\non \\
&\equiv&\, [\,\, {\bar \F}^{\bar I} ~-~ {\bar \G}^{\bar I} \fracm 1{w}
~+~
{\breve{\cal A}} {}^{\bar I} (w) ~]  ~~~.
\label{exp}
\eea
The polar multiplet is in a sense the $n = \infty$ limit of the O($2n$)
projective multiplets.  It is distinguished from them because, as
mentioned in~\cite{gkCNM}, the polar multiplet realizes a certain
$U(1)$ symmetry that cannot occur in the O($2n$) projective multiplets.
We have proposed that a minimal $N{\,=\,}2$ supersymmetric extension be
of
the form
\bea
S_{\s}[\U, \breve{\U}]
&=& \int {\rm d}^8 z \, \Big\{ ~ \frac{1}{2\pi
{\rm i}} \, \oint \frac{{\rm d}w}{w}
\cK \big( \U(w) ,
\breve{\U} (w)\, \big)  ~ \Big\}  \non \\
 &=& \int {\rm d}^8 z \, \Big\{ ~ \frac{1}{2\pi
{\rm i}} \,   \oint   \frac{{\rm d}w}{w} {}  \exp[\, ({\cal A}^I  \, +
\,
w \G^I ) \,\pa_I \,\,+\, \, ( {\breve{\cal A}}{}^{\bar I} \,-\, \fracm
1{w}{\bar \G}{}^{\bar I} ) \,{\Bar \pa}{}_{\bar I} \,  ]~ \cK \big( \F ,
{\bar \F} \, \big)  ~ \Big\} \non \\
\label{nact2} \eea
to describe a 4D, $N{\,=\,}2$ nonlinear $\s$-model. There was nothing in
our
previous discussion that prevents the K\"ahler potential~(\ref{nact2})
from taking the form given in~(\ref{kahler}).  Since the potential has a
special form, it follows that
\bea
S_{\s}[\U, \breve{\U}] &=& \int {\rm d}^8 z \, \Big\{  \, \frac{1}{2\pi
{\rm i}} \, \oint \frac{{\rm d}w}{w} \, [\,
\U^I (w) {\bar F}_{\bar I} (\breve{\U} (w)) +
\breve{\U}^{\bar I}(w)  F_I (\U (w)) \,] \Big\}
\non \\
&=& \int {\rm d}^8 z \, \Big\{ \, \frac{1}{2\pi {\rm i}} \, \oint
\frac{{
\rm d}w}{w} \, [\, \F^I \,+\, \G^I w \,+\, {\cal A}^I \,] \, \exp[\, (
\,
{\breve{\cal A}}^{\bar K} \,-\, \fracm 1{w} {\bar \G}{}^{\bar K} )\,
{\Bar
\pa}{}_{\bar K} \,  ] \, {\bar F}{}_{\bar I} ({\bar \F})   \non \\
 &{~}& {~~~~~~~~~~~~~~~~~~~~~~~~~~}~+~ {\rm {h.\,c.}}  ~
\Big\} ~~~.
\label{nact3} \eea
So at least formally, it is possible to utilize the polar multiplet to
define a $c$-map that connects a solely holomorphic function $F$ to a
4D,
$N{\,=\,}2$ supersymmetric nonlinear $\s$-model and therefore a
hyper-K\"ahler model. The action~(\ref{nact3}) can be shown to look
similar to that in~(\ref{HMact}), with the differences that (i) the real
auxiliary superfield $V$ is absent; (ii) the set of complex auxiliary
superfields $U_{\ell}$ includes infinitely many representatives,
$\ell = 1,2, \dots, \infty$; (iii) the function
${\cal P}( \G ,  {\bar
\G}, {U}_{\ell},  {\bar U}_{\ell}) $
becomes transcendental.

A specific feature of the model~(\ref{nact2}) and its special
version~(\ref{nact3}) we note, is a rigid U(1) symmetry defined by
\be
\U^I  (w)~ \longrightarrow ~\U^I  ({\rm e}^{{\rm i} \a} w)~.
\ee
This symmetry can be treated as a formal limit,  $U(1) = {\rm lim}_{n
\to \infty} \bZ_{2n}$, of the discrete symmetry~(\ref{z2n}) in the O($2n$)
multiplet model.  Since the models~(\ref{O2NAT}) and~(\ref{nact2})
possess different symmetries, it is natural to expect that these models
lead to different hyper-K\"ahler structures.

We would be remiss if we did not mention that this form of the polar
multiplet, since it depends also only on the holomorphic function $F$,
can be combined with the KVM action~\cite{KVM} by identifying the two
holomorphic functions to offer a second possible starting point for the
construction of a hyper-K\"ahlerian Vector Multiplet (HKVM) action. It
also worthwhile to note that such a HKVM model might play a role for
the effective action of $N$ = 4 supersymmetric Yang-Mills theory just as
the KVM model plays the critical role of encoding the effective action
of $N$ = 2 supersymmetric Yang-Mills theory.

\sect{Discussion}
\? The models discussed in this note appear to be avatars for a number of
interesting phenomena, 
most of which follow from the nice structure of
projective superspace~\cite{KLR}. We now discuss some of them in
turn.

\subsection{Some General Field-Space Issues}
\? All the $N{\,=\,}2$ multiplets considered in this note take the form
of
an order-$2n$ polynomial in the complex variable $w$ with (effectively)
$N{\,=\,}1$ superfield coefficients (suppressing the $I$ superscripts):
\be
 \S(w) = \sum_{k=0}^{2n} w^k\S_k~~~.
\label{GenSigma}
\ee

The terminal superfields in the series, $\S_0,\S_{2n}$, are chiral and
antichiral, respectively, and the adjacent ones, $\S_1,\S_{2n-1}$ are
complex (anti)linear superfields. In addition, all  but the $n\to\infty$
limit of~(\ref{GenSigma}) are required to satisfy the twisted reality
condition
\be
 \breve\S(w)~=~\S(w)~~~,\qquad
 \breve\S(w)~\buildrel{\rm def}\over=~
  (-1)^nw^{2n}\sum_{k=0}^{2n} \big({-1\over w}\big)^k\bar\S_k~~~.
\label{real}
\ee
This forces the `middle' superfield, $\S_n$ to be real, and so for
$n{\,=\,}1$, $\S_1$ is a {\it real\/}, not complex, linear superfield.

The first superfields $\S_0=\Phi$, are identified as the `coordinate
superfields' since their lowest components map the 4D spacetime into the
target manifold ${\cal M}_d$ and serve as local coordinates. The next
superfields, $\S_1=\G$, may be identified with tangent vectors (much as
the fermions within the coordinate chiral superfields $\Phi$ are), so
that the pair $\Phi,\G$ provides a local (super)coordinate chart for the
total space of the holomorphic (real in the $n{\,=\,}1$ case) tangent
space $T_{\cal M}$.  The interpretation that complex linear superfields
most naturally are associated with K\" ahler manifold tangent vectors
was first given in~\cite{STRG1} and provided a solution to the puzzling
fact that the component form of nonlinear $\s$-models defined in terms of
$\G$ possess~\cite{BBG} a very different form from those defined in
terms of $\Phi$.

The remaining $2n{-}3$ superfields $\S_k$, $k=2,\ldots,2n{-}2$ are
auxiliary, since their equations of motion~(\ref{surface}) are purely
algebraic. Generally (and for $n>1$), their geometrical interpretation
is as follows. The total space, $Y$, coordinatized by
$\S_0,\ldots,\S_{2n}$ enlarges $T_{\cal M}$ considerably by the
introduction of the $2n{-}3$ auxiliary superfields; in fact, ${\rm
dim}_{\bf R}(Y)=(2n+1)d$. The $(2n{-}3)d$ {\it algebraic\/}
equations~(\ref{surface}) then bring us back to the complex
$2d$-dimensional total space of $T_{\cal M}$, embedded now in the much
bigger $Y$ as an algebraic variety!

This is not dissimilar to the approach of~\cite{LG,LG2}, where
(super)string compactifications on Calabi-Yau
manifolds~\cite{CICY,Beast}
were discussed with the aid of the non-linear $\s$-model
\bea
S_{CICY} &=&\, S_{kin.} + S_{con.}       \non \\
S_{kin.} &=&\, \sum_{r=1}^m \int {\rm d}^8 z ~ w^r \, {K}^{(r)}
(\Phi,\bar \Phi) ~~~,~~~ {K}^{(r)}(\Phi,\bar \Phi) ~=~ \log\Big(
\sum_{\m_r=0}^{n_r} \left|{\Phi}_{(r)}^{\m_r}\right|^2  \Big) ~~~,
\non \\
S_{con.} &=&\, \Big[~ -{\rm i} \int {\rm d}^6 z ~ \sum_{a=1}^K {\L}^a \,
P_a(\Phi)
~+~ {\rm {h.\,c.}} ~\Big] ~~~~;    \label{CYCI}
\eea
for notation see~\cite{LG2}. Here too, one starts from a larger field
space which is then dynamically constrained to the Calabi-Yau algebraic
subvariety, defined as the the common zero-set of the simultaneous
(complex) algebraic equations $P_a(\Phi){\,=\,}0$.

 However, in~(\ref{CYCI}), the algebraic constraints on the $\Phi$-space
are enforced through the introduction of the {\it additional\/} Lagrange
multiplier superfields, $\L^a$. The models considered in this note are
more frugal: the constraints~(\ref{surface}) are enforced through
varying the very same superfields which are being eliminated. In this
respect, the present situation is somewhat more similar to the
Landau-Ginzburg approach of Ref.~\cite{LG}, where the projectivity of
the kinetic term turns some of the fields into effectively auxiliary
Lagrange multipliers. Ultimately, the approaches of Refs.~\cite{LG}
and~\cite{LG2} become analytic continuations of one another, within the
context of Witten's {\it gauged\/} linear $\s$-model~\cite{Phases}, or
different gauge choices in the non-linear $\s$-model~\cite{Elusive}. In
addition, this unifying approach firmly establishes relations to toric
geometry~\cite{AGM}. Here, however, it is not clear how such a relation
could be established, since the multiplets~(\ref{GenSigma}) are all
real~(\ref{real}), {\it except when $n\to\infty$\/}, precluding the usual
coupling to U(1) gauge fields|essential for relations to toric geometry
\`a la the works in references~\cite{Phases,AGM}.

So, the algebraic system~(\ref{surface}) seeming inherently real,
it seems quite dissimilar to the system $P_a(\Phi){\,=\,}0$, produced by
varying the chiral superpotential in~(\ref{CYCI}) by the chiral $\L^a$.
Note, however, that the whole Lagrangian superdensity
 ${\cal L}(U_\ell,\bar U_\ell,V;\Phi,\bar\Phi,\G,\bar\G)$
within the square brackets in~(\ref{HMact}), and so also the real
function $\cal P$ in~(\ref{CYCI}), may be regarded as (twice) the real
part of a holomorphic function. That is,
 ${\cal L}={\bf L}+h.c.$, where {\bf L} is a holomorphic
function of the $2n{+}1$ $N{\,=\,}1$ superfields $\S_k$ {\it before\/}
the
reality condition~(\ref{real}) is imposed. From ${\bf L}(\S_k)$, we
obtain $\cal L$ by imposing the reality condition~(\ref{real}) on the
$\S_k$'s and then adding the hermitian conjugate.
 Furthermore, the functions being constrained to vanish in
Eqs.~(\ref{surface}) are real parts of holomorphic functions|the same
ones as in Eqs.~(\ref{surface}), but before the reality
condition~(\ref{real}) enforced.

The situation is roughly as follows:

\be
\matrix{ \hfill(\S_0,\S_1,\ldots,\S_{2n-1},\S_{2n})=&Y^c
         & \tooo{~~(\ref{surface})^c~~} & T^c_{\cal M} \cr
 &\cllap{(\ref{real})}\bigg\down && \bigg\down\crlap{(\ref{real})} \cr
         \hfill(\Phi,\G,\ldots,\bar\G,\bar\Phi)=&Y
         & \tooo{~~(\ref{surface})~~} & T_{\cal M} \cr}
 \label{diag}
\ee

In fact, we may assign the (formal) weights, $\deg(\S_k)=k$, so that
$Y^c$ is a complex weighted affine space. Formally, we may assign also
$\deg(w)={-}1$, so that $\deg(\S(w))=0$, whereupon the degree of
quasihomogeneity of the action~(\ref{O2NAT}), and so also of the
holomorphic function {\bf L} is $2n$. Of course, the same is obtained
upon
using the above weight assignments {\it after\/} the contour integral is
evaluated: in the action~(\ref{HMact}), with~(\ref{O(6)}).  Thus, the
variable $w$ effectively projectivizes the whole superfield system
$Y^c=(\S_0,\ldots,\S_{2n})$, and the constraints~(\ref{surface})
describe
the real part of a complete intersection of algebraic hypersurfaces
(each
defined by a quasihomogeneous holomorphic polynomial equation) in a
weighted
{\it projective} space, $P(w^{-1},\S_0,\ldots,\S_{2n})$. The actual
(still
complex) field space, $Y^c$ where $w$ has been integrated out, may be
understood as the $w{\,=\,}1$ (affine) coordinate patch; the
compactification
of $Y^c$ into $P(w^{-1},\S_0,\ldots,\S_{2n})$ is a fairly standard
method
in algebraic geometry.

For reasons that we hope will be clearer shortly, the passage from
$T_{\cal M}\hookrightarrow Y$ to $Y^c$ is reminiscent of the fact
that the cycles used in describing the moduli space of complex
structures
on Calabi-Yau weighted complete intersections turn out themselves
to be real parts of algebraic subspaces~\cite{Periods}.

Finally, as the equations~(\ref{surface}) are used to eliminate the
auxiliary superfields $U_\ell,V,\bar U_\ell$, it is easy to determine
their degrees. Translating from $\deg(\S_k)$, we have
\be
\deg(U_\ell)=\ell{+}1~~~,\qquad \deg(V)=n~~~,\qquad
\deg(\bar U_\ell)=2n-\ell{-}1~,
\ee
and therefore with ${\cal P}={\bf P}+h.c.$,
\be
\deg_{Z}\Big({\pa{\bf P}\over \pa X}\Big) ~=~
\Big\lfloor{2n-\deg(X)\over \deg(Z)}\Big\rfloor~~~,
\ee
where $X,Z$ range over $U_\ell,V,\bar U_\ell$, and
``$\lfloor~~\rfloor$''
indicates truncation to the integral part.

Given the above general observations, one may expect to obtain, after
eliminating the auxiliary superfields $U_\ell,V,\bar U_\ell$, the same
description of $T_{\cal M}$, regardless of which of the O($2n$)
projective
multiplets one used. This, however, is not true.

Consider, for example, the O(6) case. The three equations of the
system~(\ref{surface}) are of tri-degree (2,1,1), (1,1,2) and (1,1,1),
respectively, with respect to the three auxiliary fields $U,V,\bar U$.
Upon projectivization, these are simply three quasihomogeneous algebraic
equations in three variables, where the coefficients depend polynomially
and functionally on $\F,\bar\F$ and $\G,\bar\G$, respectively; the
general methods discussed in Ref.~\cite{Beast} apply, although one must
take into account  that the coefficient functions are all (real parts of)
derivatives of a single function. In particular, as a special case of
B\'ezout's  theorem, the system~(\ref{surface}) is expected to have a
4-fold (degenerate) solution: the $\pa{\cal P}\over\pa V$ equation is
linear in $V$ and yields a unique solution for $V$ in terms of $U,\bar U$
(and of course, $\F,\bar\F,\G,\bar\G$). But then, the second equation is
quadratic in $\bar U$ and linear in $U$, so that a substitution into the
first one yields a quartic equation for $\bar U$, in terms of $\F,\,
\bar\F,\, \G,\, \bar\G$. The four solutions are related by a $\F,\,
\bar\F,\,\G,\,\bar\G$-dependent ${\bf Z}_4$ action, ${\cal Z}_4$. So,
upon elimination of the auxiliary fields $U,V,\bar U$, the O(6) model
describes four ${\cal Z}_4$-related copies of $T_{\cal M}$. Furthermore,
as ${\cal Z}_4$ is a $\F,\bar\F,\G,\bar\G$-dependent ${\bf Z}_4$-action,
the multiple copies may coincide at special subspace, ${\cal B}\subset
T_{\cal M}$.  That is, the target space of the O(6) model is ${\cal
C}^4_{\cal B}(T_{\cal M})$, a 4-fold cover of $T_{\cal M}$ branched over
${\cal B}\subset T_{\cal M}$! The case of no branching is included, in
which case ${\cal B}=\emptyset$. (For more information about branched
coverings, see Ref.~\cite{Beast} and the references therein.)

By contrast, the O(4) model has no such degeneracy, since the only
auxiliary superfield, $V$, appears linearly in its equation of motion,
${\pa{\cal P}\over\pa V}=0$, and is eliminated uniquely. On the other
hand, even without explicit calculations, it is clear that for $n>3$
the target space of the O($2n$) models will become an even higher
(branched) cover of $T_{\cal M}$.

Altogether another, but related, issue is that of dualizing the complex
(real when $n{\,=\,}1$) linear superfields $\G$ into $\J$. The $\J$'s 
being dual to the $\G$, the resulting model, in terms of $\F,\J$ (and
their conjugates), has $T^*_{\cal M}$ as the target space. In passing
from the action~(\ref{N2TMact}) to~(\ref{cmap2}), one solves
Eq.~(\ref{ltr}) which for $n{\,=\,}1$ is linear in $\G$ (and obviously
linear in $\J$). This ensures a 1--1 mapping $\G\leftrightarrow\J$, and
so also of the respective models' target spaces $T_{\cal M}\leftrightarrow
T^*_{\cal M}$.

Consider now what happens when $n{\,>\,}1$. For $n{\,=\,}2$,
Eq.~(\ref{ltr}) is cubic in $\G,\bar\G$, and so assigns three values of
$\G$ for every value of $\J$, in a $\F,\bar\F$-dependent fashion. That
is, in solving~(\ref{ltr}) for $\G,\bar\G$ in terms of $\J,\bar\J$, we
obtain three Riemann sheets for each $\J$. They all coincide at least at
$\G,\bar\G=0=\J,\bar\J$; other coincidences may appear depending on the
Lagrangian ${\cal L}(\F,\bar\F,\G,\bar\G)$, and so ultimately depending
on the choice of the function $F(\S(w))$ in~(\ref{o(4)act}); the loci of
such coincidences introduce branching. Since the $\F,\bar\F$-dependent
discrete `jump' from one to another Riemann sheet is not an isometry of
the hyper-K\"ahler structure (and which we wish to preserve), in
solving~(\ref{ltr}) we must not identify them (akin to orbifolding) but
must include them all. With the three $\F,\bar\F$-dependent $\J$-Riemann
sheets, the $\F,\bar\F,\J,\bar\J$-model's target space is then a
(possibly branched) triple cover of the dual of the $\F,\bar\F,
\G,\bar\G$-model's target space. Now, as the $\J$- and the
$\G$-spaces are contractible (as (co)tangent spaces to $\cal M$), it may
be possible to isolate any one of the $\J$-Riemann
sheets\footnote{Amusingly, this logical possibility arises because of
the real projection~(\ref{real}).}. If so, we would obtain 3 a priori
distinct $\F,\bar\F,\J,\bar\J$-models, the target space of each being the
dual of the $\F,\bar\F,\G,\bar\G$-model's target space.

For the record, the dualizing map~(\ref{ltr}) for the O($2n$) model is
\be
 \{\J,\bar\J\}
 \froto{~1\>\hbox{\scriptsize|}\>(2n{-}1)~}
 \{\G,\bar\G\}~~~,
\ee
implying that there are $2n{-}1$ $\J$-Riemann sheets. This leads to a
$\F,\bar\F,\J,\bar\J$-model the target space of whihc is a $2n{-}1$-fold
(branched) cover of the dual of the $\F,\bar\F,\G,\bar\G$-model's target
space, or to $2n{-}1$ a priori distinct dualizations of the
$\F,\bar\F,\G,\bar\G$-model, using the isolated various $\J$-Riemann
sheets. In the O(6) case, the target space of the latter we showed was,
${\cal C}_{\cal B}^4(T_{\cal M})$, a (possibly) branched 4-fold cover of
$T_{\cal M}$. Eq.~(\ref{ltr}) the produces either a
$\F,\bar\F,\J,\bar\J$-model with target space
${\cal C}^5_{\cal B'}({\cal C}_{\cal B}^4(T^*_{\cal M}))$, or 5 a priori
distinctly dualized, $n{\,=\,}2$ $\F,\bar\F,\J,\bar\J$-models with target
space ${\cal C}_{\cal B}^4(T^*_{\cal M})$, a single $\J$-Riemann sheets
can be isolated.

Notice that both the system~(\ref{surface}) and the dualizing
map~(\ref{ltr}) are {\it algebraic}. This ensures that the above
multiplicity counting is correct, except for possible degenerations and
loss of some of the solutions owing to the real nature of the
${\pa{\cal P}\over\pa V}=0$ equation, and the non-holomorphy
of~(\ref{ltr}). These can be better controlled in the complex category,
{\it i.e.}, working in the upper row of the diagram~(\ref{diag}). Of
course, {\it explicit} solutions will be possible only for rather special
choices of the function $F(\S(w))$, as noted above.

Finally, much of the relations between (super)field theory and geometry
stems from a cohomological interpretation of the massless sector of the
Hilbert space. In the finite-$n$ models considered here, we note that
the discrete symmetry~(\ref{z2n}) produces (super)sectors in the Hilbert
space and so induces (super)selection rules.

The $N{\,=\,}1$ `component' superfields $\S_k$ have distinct
transformations with respect to the ${\bf Z}_{2n}$ specified
in~(\ref{z2n}): $\F,\bar\F$ are invariant, while $\G,\bar\G$ transform
with $e^{\pm{\rm i}\a}$. Correlation functions must have a total
`$\a$-charge' of $0~({\rm mod}\,2n)$, which becomes strictly 0 in the
$n\to\infty$ limit. This symmetry may be further twisted by $N{\,=\,}1$
supersymmetric $R$-symmetries, complicating the selection rules
accordingly.

\subsection{Some Specific Field-Space Issues}
\? While the above discussion, appropriately generalized, applies even to
$N{\,=\,}2$ models with more general Lagrangians, some of the features we
wish to discuss depend exclusively on the special geometry determined by
the single holomorphic function $F(\S(w))$.

Above, we have drawn some analogies between the present models and the
Calabi-Yau models~(\ref{CYCI}) and their Landau-Ginzburg
relatives~\cite{LG}. Taking these seriously implies a novel concept for
Calabi-Yau manifolds: that they admit a special geometry metric, perhaps
even one where $K^{(r)}(\F,\bar\F)$ and $P_a(\F)$ in~(\ref{CYCI}) are all
determined by a single holomorphic function. Whether this is in any
(useful) way related to the `standard' Einstein-K\"ahler-Yau metric
remains an open question, but seems well worth a detailed study, since
there is only an infinite iteration procedure for constructing the
`standard' metric. Moreover, the physically relevant metric receives
further world-sheet perturbation and instanton corrections. While it may
seem nothing short of a miracle, it is logically possible that the
special geometry metric, suggested by this $N{\,=\,}2$ approach and with
a suitable choice of $F(\S(w))$ may shortcut this doubly infinite
iterative procedure and produce a useful metric on Calabi-Yau manifolds.
Furthermore, as we discussed, the dualizing map~(\ref{ltr}) is a {\it
multi-valued} map  $T_{\cal M} \leftrightarrow T^*_{\cal M}$. Now,
variations of the complex structure of a Calabi-Yau manifold are
parametrized by $T_{\cal M}$-valued 1-forms, while variations of the
complexified K\"ahler class are parametrized by $T^*_{\cal M}$-valued
1-forms. The dualizing map~(\ref{ltr}) then naturally maps between these
two types of variations, and therefore also between the two corresponding
moduli spaces. However, as we have shown above, this map is
{\it multi-valued\/}, and this must be taken into account.

On the other hand, the special geometry is naturally given on the {\it
moduli\/} space of Calabi-Yau manifolds~\cite{Rolling,Andy}. As
discussed above, the target space of O($2n$) models is closely related to
$T_{\cal  M}$, and we can now take $\cal M$ to be the {\it complex
structure  moduli space\/} for a Calabi-Yau manifold rather than the
manifold  itself. As Ref.~\cite{Andy} shows, special geometry is very
closely related  to the so-called Hodge fibration over a complex moduli
space $\cal M$ of a larger space containing $T_{\cal M}$, the holomorphic
tangent space to $\cal M$|indeed very similar to our situation as
described above. The special geometry metric turns out to be remarkably
well determined in terms of the periods of the holomorphic volume
form~\cite{Rolling,Andy}, for which a natural choice of cycles includes
the real parts of some straightforward algebraic subspaces of the
Calabi-Yau manifold~\cite{Periods}. As alluded to above, this is very
reminiscent of the r\^ole of the reality `projection' as shown in the
diagram~(\ref{diag}). How far these parallels can be exploited further
remains another open question.

Finally, if the chiral superfields are valued in the adjoint
representation of some compact Lie algebra which possesses a number of
non-trivial Casimir invariants, then one obvious choice for the
$F$-functions is given by a (non)linear combination of such invariants.
Remarkably, this again seems to draw parallels with the description of
Calabi-Yau complete intersection manifolds~\cite{LG2}. There, the sought
after manifolds were embedded into products of complex projective
spaces, which in turn may be represented as cosets $\bC
P^n=SU(n{+}1)/U(n)$.  Whether a suitable generalization of the standard
coset construction, and furthermore subject to algebraic constraints to
describe embedded complete intersections is possible with the $N{\,=\,}2$
multiplets used here remains an open but intriguing question, ultimately
leading to an $N{\,=\,}2$ (and non-abelian) generalization of the linear
$\s$-model \`a la Witten~\cite{Phases}.

${~~~}$ \newline
${}~~~~~~~~~~~~~~~~~~~~~~$ \
\parbox{5.3in}{\raggedright
{\it ``Originality does not consist in saying what no
      one has ever said before, but in saying exactly
       what you think yourself.''\/}  \\
${}~~~~~~~~~~~~~~\qquad \qquad \qquad
\qquad \qquad \qquad \qquad ~~~~~{}$   James Fitzjames Stephen}

$${~~~}$$
\noindent
{\bf {Acknowledgments}} \newline \noindent
The authors wish to acknowledge very helpful discussions with
S. Theisen. The work of SMK was also supported in part from
RFBR-DFG grant, project No 96-02-00180; INTAS
grant, INTAS-96-0308; Grant center for natural
sciences of the Ministry of General and
Professional Education of Russian Federation,
project  No 97-6.2-34.


\end{document}